\documentclass[runningheads]{llncs}

\usepackage{cite}
\usepackage{amsmath,amssymb,amsfonts}
\usepackage{graphicx}
\usepackage{textcomp}
\usepackage{xcolor}

\usepackage{booktabs} 
\usepackage{amsmath,amssymb}
\usepackage{array}
\usepackage{mdwtab}
\usepackage{subcaption} 
\usepackage{fixltx2e}
\usepackage{dblfloatfix} 
\usepackage{url}
\usepackage{xspace}
\usepackage{numprint}
\usepackage{algorithm}
\usepackage{algorithmicx}
\usepackage{algpseudocode}
\usepackage{float}
\usepackage{placeins}
\newfloat{algorithm}{t}{lop}
\usepackage{flushend} 
\usepackage{todonotes}
\usepackage{cleveref}

\usepackage[inline]{enumitem}

\newcommand{\Oh}{\ensuremath{\mathcal{O}}}

\newcommand{\myvec}[1]{\mathbf{#1}}
\newcommand{\vent}[2]{\myvec{#1}[#2]}
\newcommand{\size}[1]{\operatorname{size}(#1)}

\newcommand{\ie}{\textit{i.\,e.}\xspace}

\newcommand{\eg}{\textit{e.\,g.}\xspace}

\newcommand{\etal}{\textit{et al.}\xspace}

\newcommand{\etc}{etc.\xspace}

\newcommand{\parmetis}{\textsc{ParMETIS}\xspace}
\newcommand{\libtopomap}{\textsc{LibTopoMap}\xspace}

\newcommand{\kahip}{\textsc{KaHIP}\xspace}
\newcommand{\mapkahip}{\textsc{KaHIP\_map}\xspace}
\newcommand{\parhip}{\textsc{ParHIP}\xspace}
\newcommand{\mpipp}{\textsc{MpiPP}\xspace}
\newcommand{\jostle}{\textsc{Jostle}\xspace}

\newcommand{\scotch}{\textsc{Scotch}\xspace}
\newcommand{\ptscotch}{\textsc{PT-Scotch}\xspace}

\newcommand{\topomatch}{\textsc{TopoMatch}\xspace}
\newcommand{\xpulp}{\textsc{xtraPulp}\xspace}

\newcommand{\argmin}{\operatorname{argmin}\xspace}

\newcommand{\cc}{\operatorname{Coco}\xspace}

\newcommand{\ouralgo}{\textsc{ParHIP\_map}\xspace }

\newcommand{\real}{\textsc{real}\xspace}
\newcommand{\ba}{\textsc{BA}\xspace}
\newcommand{\rhg}{\textsc{RHG}\xspace}
\newcommand{\rmat}{\textsc{Rmat}\xspace}

\usepackage{ifthen}
\newboolean{confversion}
\setboolean{confversion}{true}

\newcommand{\budget}[1]{{\color{blue}Page budget: #1}}
\renewcommand{\budget}[1]{}

\graphicspath{{./figures/}}

\def\BibTeX{{\rm B\kern-.05em{\sc i\kern-.025em b}\kern-.08em
    T\kern-.1667em\lower.7ex\hbox{E}\kern-.125emX}}
    
\begin{document}

\title{An MPI-based Algorithm for Mapping Complex Networks onto Hierarchical Architectures
\thanks{This work is partially supported by German Research Foundation (DFG) grant ME 3619/3-2
(FINCA) within Priority Programme 1736 and by DFG grant ME 3619/4-1 (ALMACOM) as well as by the Austrian Science Fund~(FWF, project P 31763-N31).
}
}

\titlerunning{An MPI-based Mapping Algorithm for Hierarchical Architectures}
%
\author{Maria Predari\inst{1} \and
 Charilaos Tzovas\inst{1} \and
Christian Schulz\inst{2} \and \\
Henning Meyerhenke\inst{1}}
\authorrunning{M. Predari et al.}
%
\institute{Humboldt-Universit\"at zu Berlin, Berlin, Germany\\
\email{\{charilat,predarim,meyerhenke\}@hu-berlin.de}
\and
Universit\"at Heidelberg, Heidelberg, Germany\\
\email{christian.schulz@informatik.uni-heidelberg.de}
}

\maketitle


\begin{abstract}
Processing massive application graphs on distributed memory systems requires
to map the graphs onto the system's processing elements (PEs). This task
becomes all the more important when PEs have non-uniform communication costs
or the input is highly irregular.
Typically, mapping is addressed using partitioning, in a two-step approach 
or an integrated one. Parallel partitioning tools do exist; yet,
corresponding mapping algorithms or their public implementations all have major 
sequential parts or other severe scaling limitations.

In this paper, we propose a parallel algorithm that maps graphs
onto the PEs of a hierarchical system.
Our solution integrates partitioning and mapping; it models the system hierarchy
in a concise way as an implicit labeled tree. 
The vertices of the application graph are labeled as well, and these vertex labels induce the mapping.
The mapping optimization follows the basic idea of parallel label propagation,
but we tailor the gain computations of label changes
to quickly account for the induced communication costs.
Our MPI-based code is the first public implementation of a parallel graph mapping algorithm;
to this end, we extend the partitioning library \parhip. 
To evaluate our algorithm's implementation, we perform comparative experiments
with complex networks in the million- and billion-scale range.
In general our mapping tool shows good scalability on up to a
few thousand PEs. Compared to other MPI-based competitors, our algorithm
achieves the best speed to quality trade-off and our quality results are even better
than non-parallel mapping tools.

\end{abstract}


\section{Introduction}
\label{sec:intro}

\budget{1}

Task mapping is the process of assigning tasks of a parallel application
onto a number of available processing elements (PEs)
and is an important step in high-performance computing.
One reason for the importance of mapping is non-uniform memory access (NUMA),
common in many modern architectures, where PEs close to each other
communicate faster than PEs further away.
The importance stems from the fact that communication
is orders of magnitude slower than computation.
To alleviate those issues, task mapping is often
used as a preprocessing step.
Successful mapping solutions assign pairs of heavily-communicating
tasks ``close to each other'' in the parallel system, so that
their communication overhead is reduced.
%
%
Moreover, the network topologies of parallel architetures
exhibit special properties that can be exploited during mapping.
A common property is that PEs are
hierarchically organized into, e. g., islands, racks, nodes,
processors, cores with corresponding communication links of similar
quality.

Furthermore, mapping becomes even more important when the application's structure and communication 
pattern are highly irregular. While partitioning the application 
may work well for mesh-based numerical simulations,
large graphs derived from social 
and other complex networks pose additional challenges~\cite{xtrapulp}.
Typical examples are power-law~\cite{Barabasi-power-law}
and small-world graphs~\cite{kleinberg-small-word}.
The former are characterized by highly skewed vertex degree distribution,
the latter exhibit a particularly low graph diameter.
Distributed graph processing systems, such as Apache Giraph
and GraphLab~\cite{GraphLab}, are made to run analytics on such graphs.
For some algorithms, in particular those with local data access, they report good scaling
results~\cite{GraphLab}.
For non-local or otherwise complex analytics and highly irregular inputs,
running times and scalability of these systems can become unsatisfactory~\cite{ilprints1039}.
%
%
Consequently, 
designing MPI-based graph processing
applications is necessary to scale to massive instances with high performance.
Considering the above and the number of PEs in modern architectures
(a number expected to increase in the future),
task mapping can have significant impact in the overall application 
performance~\cite{10.1007/978-3-642-32820-6_82,10.1145/1542275.1542295}.
Good mapping algorithms should be able to improve the quality of the final mapping, and
additionally, they need to be fast in order not
to degrade the overall performance of the application.
Thus, developing MPI-based mapping algorithms with good scaling
behavior becomes all the more crucial.
Since finding optimal topology mappings is NP-hard~\cite{hoefler-topomap},
heuristics are often used to obtain fairly good solutions
within reasonable time~\cite{5713190, 1676925}.  

Our contribution is an MPI-based, integrated mapping algorithm
for hierarchically organized architectures, implemented within \parhip.
To model the system hierarchy and the corresponding
communication costs, we use an implicit bit-label representation,
which is very concise and effective.
Our algorithm, called \ouralgo, uses parallel label propagation
to stir the mapping optimization.
As far as we know \ouralgo is currently the only
available implementation for parallel mapping. Experiments
show that it offers the best speed to quality trade-off; having on average
62\% higher quality than the second best competitor (\parhip), and being
only 18\% slower than the fastest competitor (\xpulp), which favors speed over quality.
Moreover, our algorithm scales well up to \numprint{3072} PEs
and is able to handle graphs of billion edges, with the least failing rate
among other MPI-based tools due to memory or timeout issues on massive complex networks.
Moreover, compared to a sequential baseline mapping algorithm, \ouralgo has on average
10\% better quality results.

\section{Background}
\label{sec:back}

\budget{2}


We model the underlying parallel application with a graph $G_a = (V_a, E_a, \omega_a)$,
where vertex $u_a$ represents a computational task of the application
and an edge $e_a$ indicates data exchanges between tasks.
The amount of exchanged data is quantified by the edge weight $\omega_a(e_a)$.
Network information for hierarchically organized systems is often
modeled with trees\footnote{In a tree topology, leaf vertices correspond to PEs, internal nodes to switches.}~\cite{Schulz2017a, Glantz18};
we do so as well, but implicitly (see Sec.~\ref{sec:algo}).
The bottom-up input description of the topology follows \kahip:
$H = \{h_0,h_1,\ldots, h_{l-1}\}$ denotes the number of children of a 
node per level, where $l$ is the number of hierarchy levels;
\ie, each processor has $h_0$ cores,
each node $h_1$ processors, and so on.
We also define the set of PEs as $V_p$ of size $k =\prod_{i=0}^{l-1} h_i$.
Communication costs are defined via $D= \{d_0, d_1, \ldots, d_{l-1}\}$ such
that PEs with a common ancestor in level $i$ of the hierarchy 
communicate with cost $d_i$; \ie, two cores in the same
processor communicate with $d_0$ cost,
two cores in the same node but in different processors with $d_1$,
and so on.
We use $d(u_p, v_p)$ to indicate the time for one
data unit exchange between PEs $u_p$ and $v_p$ (\ie, their distance).

A $k$-way partition of G divides $V$ into $k$ blocks $V_1,\ldots, V_k$,
such that $ V_1 \cup \ldots \cup V_k = V,~V_i \neq \emptyset$ for $i = 1,\ldots, k$ and $V_i\cap V_j = \emptyset$ for $i \neq j$.
Graph partitioning aims at finding a $k$-way partition of G that optimizes an objective function
while adhering to a balance constraint. The balance constraint demands that the sum of node weights in each block does
not exceed a given imbalance threshold $\epsilon$.
Moreover, the objective function is often taken to be the edge-cut of the partition $\sum_{i<j} w(E_{ij})$,
where $E_{ij} := \{\{u, v\} \in E : u \in V_i, v \in V_j\}$.

A mapping, $\mu: V_a \mapsto V_p$, is defined as a nearly balanced assignment of
computational tasks onto PEs such that, for some imbalance parameter $\varepsilon \geq 0$,
$ \vert \mu^{-1}(v_p) \vert \leq (1+\varepsilon)\cdot \lceil \vert V_a \vert / \vert \mu(V_a) \vert
  \rceil $
for all $v_p \in \mu(V_a)$. Hence, $\mu(\cdot)$ induces a balanced
partition of $G_a$ with blocks $\mu^{-1}(v_p)$, $v_p \in V_p$.
Or inversely, a mapping $\mu$ defines a one-to-one mapping
from $k$ balanced blocks of $V_a$ to $k$ PEs of $V_p$.
%
%
To steer an optimization process for obtaining good mappings,
different objective functions have been proposed~\cite{hoefler-topomap}.\footnote{
Most theoretical metrics can only
approximate the communication overhead of
the application since communication during real-time execution
can be affected by many external factors
(e.g., network traffic and overhead from competing jobs).
}
A widely used~\cite{Pellegrini11static} mapping objective
is $\cc(\cdot)$ (also referred to as \emph{hop-byte} 
or \emph{qap}), defined as:
\begin{align}
\label{eq:objFuncCommCost2}
\cc(\mu) &:= \sum_{\substack{e_a \in E_a\\{\mbox{\tiny $e_a = \{u_a, v_a\}$}}}} \omega_a(e_a)~d(\mu(u_a), \mu(v_a))
\end{align}
%
Intuitively speaking, placing pairs of highly
communicating tasks in nearby PEs minimizes $\cc(\cdot)$.

\subsection{Related Work}
\label{sec:rel}

In this section we focus on related techniques
for parallel graph partitioning and sequential and parallel
task mapping. For more details we refer the reader to the overview articles
for task mapping~\cite{Pellegrini11static,hoeflerSurvey} and
for graph partitioning~\cite{SPPGPOverviewPaper}.

\paragraph{Parallel graph partitioning.}
Graph partitioning is closely related to task mapping. First, because
it is often used as a building block for mapping and second,
because it substitutes mapping when no network information is available.
A trivial mapping can be computed from the solution of
a graph partitioner, simply by assigning block $i$ to PE $i$ (identity mapping).
To this end, a graph-based metric, such as the edgecut, \ie, the total weight
of edges between blocks, is minimized.
Some popular parallel partitioners are
\parmetis~\cite{SchloegelKK02parallel}, \parhip~\cite{DBLP:journals/tpds/MeyerhenkeSS17}, and
\ptscotch~\cite{Pellegrini12scotch}. 
These tools all follow the multilevel framework, performing one or more cycles of the following procedure:
they construct a hierarchy of successively coarser graphs, find an initial solution on the coarsest graph
and project the solution successively to the original graph, while refining it on every level.
Graph coarsening is often based on edge matching~\cite{SchloegelKK02parallel}
or label propagation~\cite{Raghavan:2007fk}, while initial partitioning
uses recursive bisection, local heuristics
or evolutionary algorithms. The main bottleneck for high scalability
in these tools seems to be the high memory usage
due to successive coarsening and the poor scaling of the
initial partitioning phase. \xpulp~\cite{xtrapulp}, a single-level parallel partitioning tool
that uses label propagation, avoids the scalability issue. This advantage comes with the price of reduced quality, though.

\paragraph{Mapping tools.}

Existing mapping algorithms are grouped into two categories:
integrated approaches and two-phase ones.
Integrated approaches address the mapping problem using the network
information directly, without decomposing the problem into independent sub-problems.
Examples of integrated approaches are included in
\scotch~\cite{Pellegrini07scotch} and \kahip~\cite{sandersschulz13}.
\scotch uses dual recursive bisection (DRB)~\cite{Pellegrini94static} to
partition both the application graph and the network
topology into two blocks recursively.
Embedded sectioning~\cite{DBLP:journals/ijhpca/KirmaniPR17} and
Recursive multisection~\cite{chan2012impact,6495451} 
follow a similar technique.
Recently, Faraj \etal~\cite{faraj_et_al:LIPIcs:2020:12078}
proposed an integrated mapping approach
that uses fast label propagation
and a more localized local search to achieve mapping solutions
of high quality.
\libtopomap~\cite{hoefler-topomap}, \topomatch~\cite{jeannot:hal-00921605}
and \mpipp~\cite{10.1145/1183401.1183451} are typical examples of the
two-phase approach, where mapping is solved in two steps. The first step
usually involves an established partitioner that obtains a balanced partition.
The second step then assigns the resulting blocks to the PEs while minimizing
a mapping objective,
\eg using a greedy approach~\cite{hoefler-topomap, Brandfass2013372}
or metaheuristics~\cite{Brandfass2013372,Schulz2017a}.

To the best of our knowledge, all current 
mapping algorithms or their public implementations have major sequential parts.
\libtopomap and \topomatch use parallel partitioning,
but the mapping step is completely sequential.
Regarding integrated approaches, \ptscotch offers parallel mapping only for
the trivial case where the underlying network topology corresponds to a complete graph,
which is simply partitioning.\footnote{In the documentation of \ptscotch (6.0.1),
 there is a comment about implementing parallel mapping algorithms
 for more target architectures in future releases.
}
Moreover, the \jostle authors briefly discuss a parallel mapping
extension of their sequential approach but do not provide enough details
nor an implementation~\cite{Walshaw-MPDD-07}.

\subsection{Parallel Label Propagation with Size Constraints}
Our mapping approach uses the parallel label propagation algorithm (LPA)
with size constraints~\cite{DBLP:journals/tpds/MeyerhenkeSS17},
as implemented in \parhip.
We discuss the algorithm and its implementation here for self-containment reasons. 
In its sequential form, LPA starts with some partition (depending on the algorithm's purpose)
and iterates over all vertices. At each vertex $v$, the block number (= label) of $v$ is set
to the dominant one in the neighborhood of $v$ (= block with highest total edge weight incident to $v$).
If a size constraint is imposed, then the
dominant block that can still host $v$ is chosen. The loop over all vertices vertices is repeated,
\eg a fixed number of times or until no more changes occur.

The parallel version is implemented as follows.
First, each PE gets a subgraph of the input graph consisting of a
contiguous range of nodes in the interval $I:= [a\ldots b]$, the edges incident to the nodes of those blocks,
as well as the end points of edges which are not in $I$ (so-called ghost or halo nodes). 
In any case, the graph data structure only stores edges incident to local vertices.
To parallelize the label propagation algorithm, each PE visits all local vertices in a random order.
A vertex $v$ is moved to the block that has the strongest eligible connection such that the block will not be overloaded.
During the course of the algorithm, local vertices can change their block and hence the blocks in which halo vertices are contained can change as well. 
Communication is expensive, so instead of updating labels of halo vertices every time they change, the
algorithm follows an overlapping scheme, organized  in phases.
A node is called interface node if it is adjacent to at least one ghost node.
The PE associated with the ghost node is called adjacent PE. Each PE stores
a separate send buffer for all adjacent PEs. During each
phase, the algorithm stores the block label of interface nodes that have
changed into the send buffer of each adjacent PE of this
node. Communication is then implemented asynchronously.
In phase $i$, the current updates are sent to the adjacent PEs 
and each PE receives the updates of the adjacent PEs
from round $i-1$, for $i > 1$.

  For maintaining the weight of blocks, the algorithm uses
  two different approaches, one for coarsening
  and another for uncoarsening.
During coarsening, the algorithm uses a localized approach for keeping up
with the block weight since the number of blocks is high and the balance constraint is not tight.
Roughly speaking, a PE maintains and
updates only the local amount of node weight of the blocks
of its local and ghost nodes.
Due to the way the label
propagation algorithm is initialized, each PE knows the
exact weights of the blocks of local nodes and ghost nodes
in the beginning. Label propagation then uses the local information to bound the block weights.
Once a node changes its block, the local block weight is updated. This
does not involve additional communication.

During uncoarsening a different approach is taken compared to coarsening since
the number of blocks is much smaller.
This bookkeeping approach is similar to the one in
\parmetis~\cite{SchloegelKK02parallel}. Initially, the exact block weights of
all $k$ blocks are computed locally. The local block weights
are then aggregated and broadcast to all PEs. Both can be
done using one allreduce operation. Now each PE knows the
global block weights of all $k$ blocks. The label propagation
algorithm then uses this information and locally updates
the weights. For each block, a PE maintains and updates the
total amount of node weight that local nodes contribute to
the block weights. Using this information, one can restore
the exact block weights with one allreduce operation which
is done at the end of each computation phase.

\section{Parallel Mapping Algorithm}
\label{sec:algo}
\ifthenelse{\boolean{confversion}}
           {\budget{4}}
           {\budget{5}}

In this section, we present the main technical contributions of the paper.
This includes an integrated mapping algorithm for distributed memory systems
based on a concise representation of the hierarchical network topology via bit-labels.
Our algorithm uses a parallel local search refinement process,
where gain computations for label changes account for communication in
the network topology.
The bit-label network representation allows a quick gain evaluation.
As far as we know, this is the first (implemented and publicly available)
parallel mapping algorithm for distributed-memory systems.

\subsection{Network Topology Representation}

To encode a tree network topology,
most representations typically store $l$ numbers for each PE, one for each tree hierarchy level. 
However, in our work we use a concise bit-label representation that has
very low space requirement.
For each vertex $ v \in V_p$ (\ie, each PE),
we only store a single number as a bit-label. This number is also
hidden within the labels of vertices in $V_a$ as a bit-prefix.
This enables us to use label propagation on $G_a$ for the refinement process and quickly
evaluate distances between PEs.

\begin{figure}[bt]
  \centering
\includegraphics[scale=0.20]{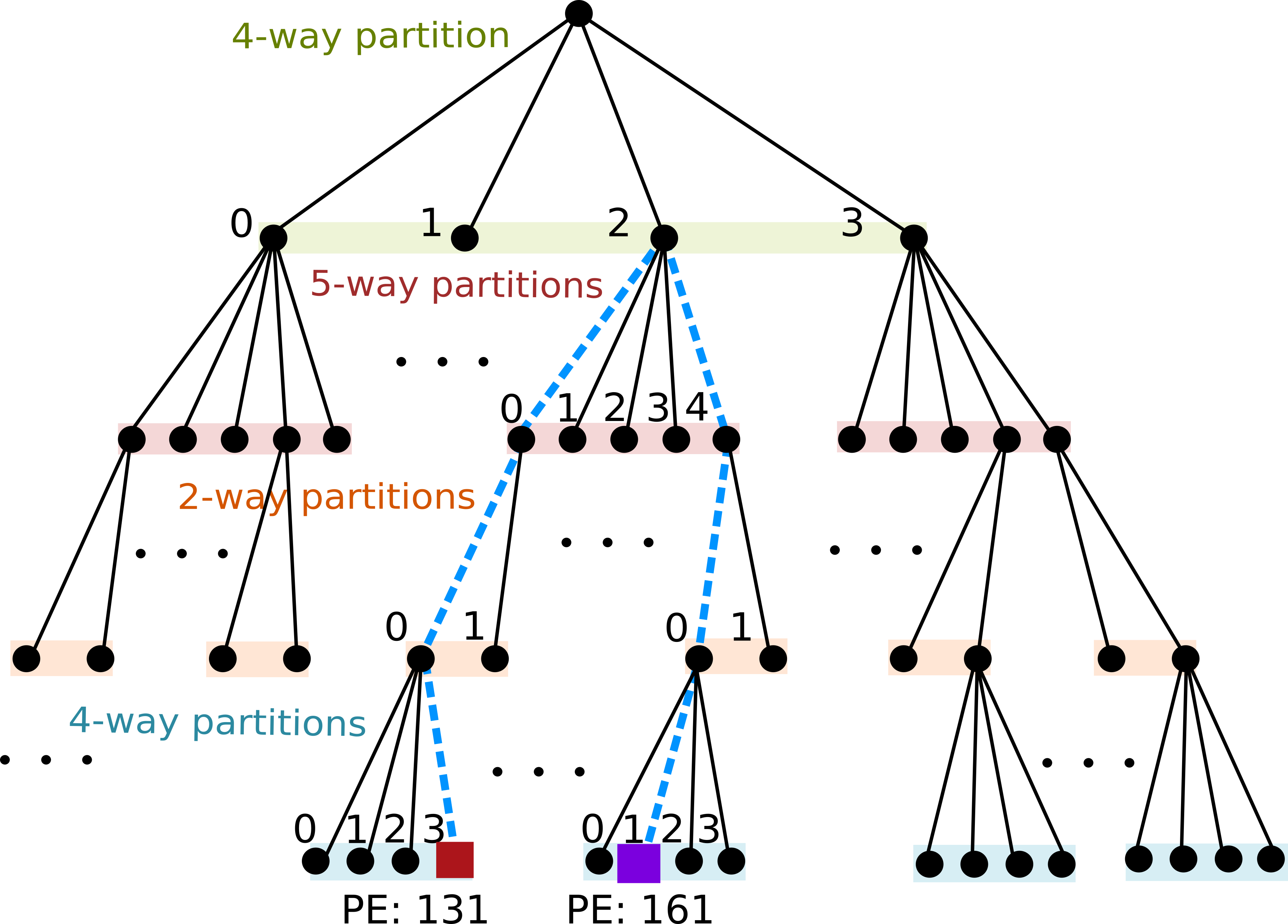} 
\caption{\label{fig:topology} Example of implicit tree topology. The system
  has $160$ PEs in one island with four racks (first level),
  five nodes per rack (second), two processors per node (third)
  and four cores per processor (forth). PEs take labels from $0$ to $231$ but not
all labels are used due to different number of children per level.}
\end{figure}

The bit-label of a given vertex $ v_p \in V_p$
encodes the full ancestry of $v_p$ in the tree.
The ancestor of $v_p$ in level $i$ can be viewed as a block
of an implicit partition in that level (see Figure~\ref{fig:topology}).
The local numberings of all ancestors/blocks of $v_p$
are encoded in the bit-label of $v_p$ and indicate
ownership of the vertex in the tree hierarchy
(in which rack, processor, node \etc, it belongs to).
The red vertex in Figure~\ref{fig:topology}
has a label of \numprint{131} ($10|000|0|11$ in binary).
Reading the bit-label from left to right, we have
that $v$ belongs to block $2$ of the first level partition,
block $0$ of the second level and so on. For the network topology this
means that PE $v$ belongs to rack $2$, local processor $0$,
local node $0$ and has local core number $3$.
To construct the labels for the available PEs we use Algorithm~\ref{alg:buildLabel}.
The bit-label of each PE is divided into $l$ sections, each containing $s[i]$ bits with $0 \leq i < l$,
as in Line~\ref{line:sectionsize}. 
Each subsection $r_i$ gives the local numbering of the ancestor node of the current
PE in level $i$.
For each PE $p$
we loop through the levels of the tree hierarchy
and encode the local numbering of each ancestor node
in the bit-label of $p$ (see Lines~\ref{line:main-for}
to~\ref{line:end-main-for}).

To retrieve the distance between two PEs, one needs to find their common ancestor in
the tree topology. To achieve this in our representation, we apply the
bit-wise operation $xor(\cdot,\cdot)$ on the two bit-labels.
We find the level of the common ancestor by
finding the section $r_i$ which contains the
leftmost non-zero bit on the result of $xor(\cdot,\cdot)$.
In the example of Figure~\ref{fig:topology}, the leftmost non-zero bit
of the squared vertices is in the second section. This corresponds
to the second level in the tree (illustrated with the blue dashed line).
The time complexity of finding the section of the
leftmost non-zero bit is $\Oh(\log l)$. Note that modern
processors often have hardware implementation of a count leading zeros operation.
This makes the identification of the leftmost non-zero bit a constant time operation.\footnote{
  This holds under the realistic assumption that for any bit-label $v_p$, the size
  $\log v_p=\Oh(\log k)$ of the binary numbers is smaller than the size of a machine word.}

\begin{algorithm}[bt]
  \begin{scriptsize}
  \begin{algorithmic}[1]

      \Function{BuildLabel}{$\myvec{H}$}
      \State \textbf{Input:} $\myvec{H}$, \ie, number of children per node for each hierarchy level
      \State \textbf{Output:} $\myvec{x}$, \ie, array of bit-label representation for PEs
      \State $l \gets \size{\myvec{H}} $
      \Comment{number of hierarchies in the tree topology}
      \State $s[i] \gets \lceil \log_2 h_i\rceil$ \label{line:sectionsize}
      \Comment{array of size $l$}
      \State $k \gets \prod_{i=1}^{l} h_i$
    \For{$p \gets 0$ to $k-1$}\label{line:for-p}    
      \State $t \gets p $ 
          \For{$i \gets 0$ to $l-1$}\label{line:main-for}
          \State $\vent{r}{i} \gets t \bmod h_i$
          \State $t \gets t / h_i$
          \State $\vent{x}{i} \gets \vent{r}{i} << (i *s[i])$
          \EndFor  \label{line:end-main-for}
     \EndFor 
     \State \textbf{return} $\myvec{x}$
    \EndFunction
    
  \end{algorithmic}
  \caption{Algorithm for building the bit-label representation for PEs.}
  \label{alg:buildLabel}
  \end{scriptsize}
\end{algorithm}


\subsection{Refinement \& Gain Computation}

\begin{figure*}
\centering
\begin{subfigure}{0.47\textwidth}
\centering
\includegraphics[scale=0.13]{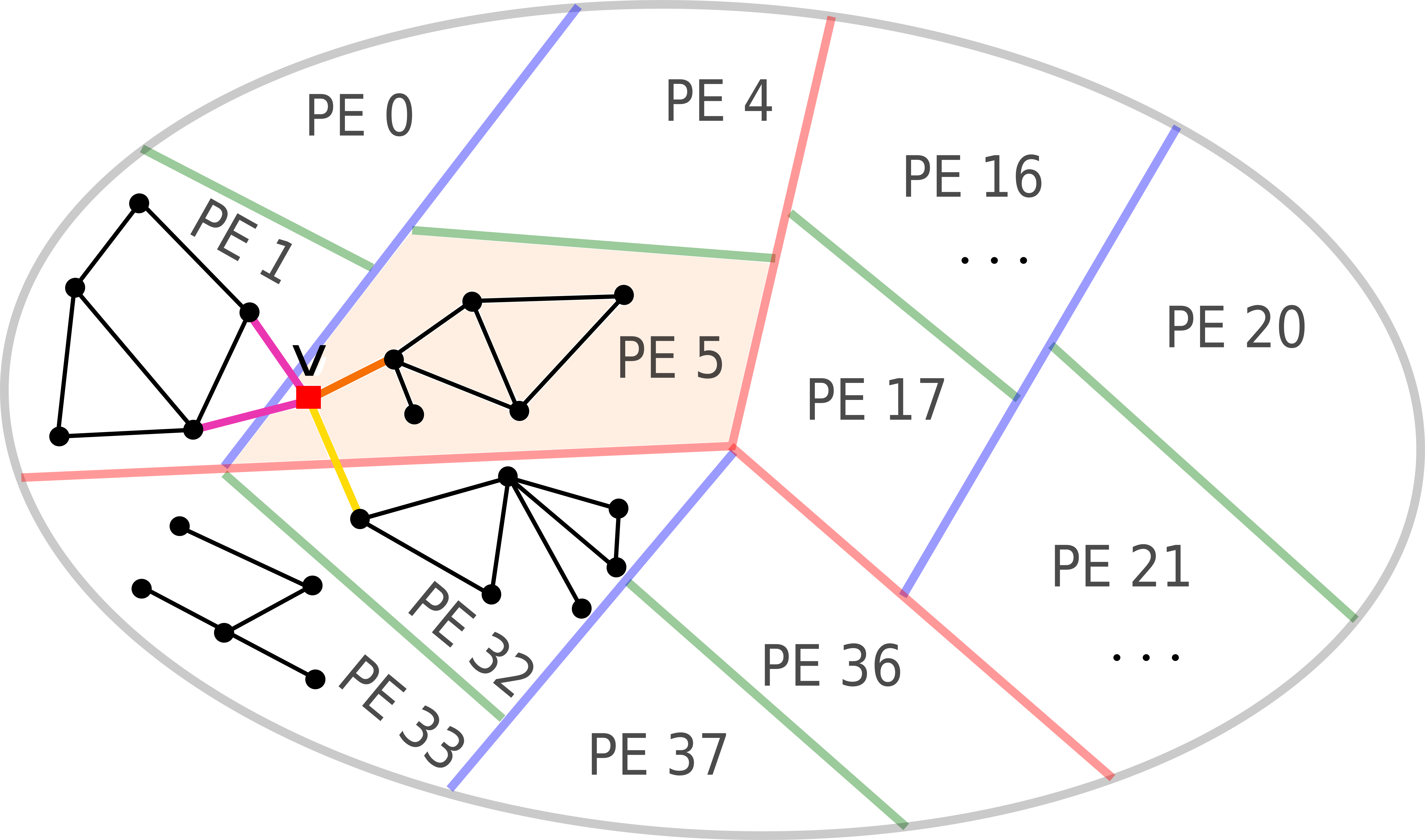} 
\caption{ \label{1a} Application graph $G_a$. }
\end{subfigure}
\begin{subfigure}{0.47\textwidth}
\centering
\includegraphics[scale=0.19]{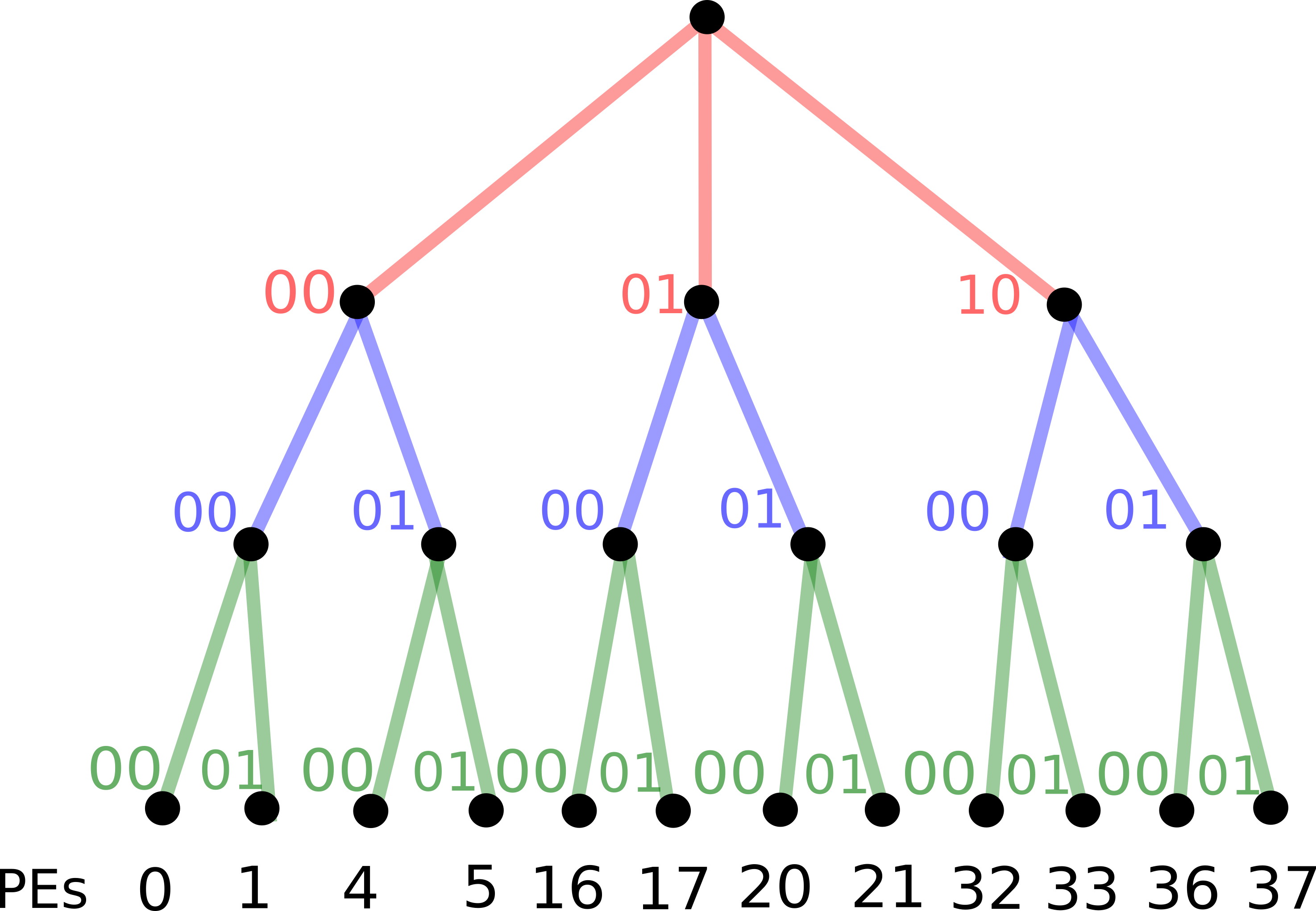} 
\caption{\label{1b} Implicit tree topology.}
\end{subfigure}
  \caption{
    \label{fig:mapping}
          {\small Mapping from $G_a$ to the (implicit) tree topology: colored lines in~(a) indicate the cuts
            induced by the tree levels in~(b). The decimal PE numbers in (a) correspond to the combined bit-labels on a path from the root to the PE leaf in (b).
          }
  }
\end{figure*}

Our parallel mapping approach is an integrated solution method that performs
one cycle of the multilevel framework. More precisely, we coarsen the graph,
compute an initial partition and uncoarsen the solution while refining with
local search based on a mapping objective such as~\Cref{eq:objFuncCommCost2}. 
In a parallel setting, each graph vertex $v_a$
has a local vertex label, within the PE, and a global one. 
The global vertex labels can be used to induce a mapping $\mu(\cdot)$ 
onto PEs via their prefixes. 
For instance, in Figure~\ref{fig:mapping} all vertices of the shaded block of $G_a$
have labels with prefix $00|01|01$, implying a mapping to PE $5$.
Using the concise network representation, retrieving communication costs and evaluating 
$\cc$ can be performed quickly for all edges of $G_a$. 
For an edge $e_a = (u_a,v_a) \in E_a$, the prefixes of $u_a$ and $v_a$
are used to compute the communication costs between $\mu(u_a)$ and 
$\mu(v_a)$, by returning the leftmost non-zero
bit on the result of $xor(u_a,v_a)$.
Overall, through the bit-label information, we can quickly
refine an initial mapping using parallel
label propagation. We use the size-constrainted version of the algorithm
and tailor the gain computations of a potential vertex move
to account for the induced communication costs.

The optimization process works as follows: 
all PEs visit their local vertices in random order and consider
moving a given vertex $v$ into another block from the set of candidates
$R(v)= \{\mu(u): u \in N(v) \} \subseteq V_p$ ($N(v)$ is the neigborhood of $v$).
The algorithm performs the move that induces the maximum improvement in $\cc$ as long as
the size constraint is respected.
To compute the best block assignment, we temporarily move $v$ to each block in $R(v)$ and calculate the communication cost
for all possible block assignments; finally, we set
$\mu(v) := \argmin_{\substack{b\in R(v)}} \left( \sum_{u\in N(v)} w(v,u) \; d_{G_p}(b, \mu(u)) \right)$.
If the maximum reduction is induced by keeping the vertex in the current block,
we do not perform any move.
In Figure~\ref{1a} and for an implicit tree distance of $D= \{2, 4, 10\}$, the maximum
reduction for vertex $v$, with label $00|01|01|***$.
After a certain number of moves, we do a global communication step to update the labels of
the halo nodes of each PE.
To handle overloaded blocks, we keep the
same modifications to the block selection rule that was proposed in~\cite{DBLP:journals/tpds/MeyerhenkeSS17}.
The process is repeated for a user-defined number of rounds.
This is repeated for each refinement level and on the original graph.

\subsection{Overall Approach}
Here we present a more detailed description of the fully parallel mapping algorithm designed for
distributed-memory systems.
We implement our algorithm in \parhip.
For the coarsening step we use the parallel size-constrained label
propagation as implemented in \parhip, without any modifications.
Each PE computes clusters of its local graph and aggregates them in super-vertices
in parallel until the coarsened graph becomes small enough.
The distributed coarse graph is then collected on each PE and is partitioned
using a min-cut/max-flow algorithm within an evolutionary framework~\cite{kaffpae}.
The best solution among all
PEs is kept and broadcast back to them. For the uncoarsening step each PE performs a local
search using the parallel size-constrained label propagation on their local part
of the coarse graph. This step is repeated for each level of the hierarchy to
refine the solution. At this point, we adapt the parallel label propagation to account
for the communication overhead among PEs. More precisely, we modify the block selection
step during vertex moving and we use the implicit topology representation to quickly
retrieve the distance costs among PEs. During gain computation we
change the objective from edgecut to $\cc(\cdot)$.

\paragraph{Avoiding memory issues.}
As already mentioned in Section~\ref{sec:rel}, classic multilevel algorithms have
high memory usage caused by multiple cycles of successive coarsening. 
Our algorithm performs only one cycle of the multilevel framework, but successive coarsening
(even in one cycle) can still damage the scalability of our approach.
Moreover, the previous implementation of \parhip (without our contributions for parallel mapping)
uses block partitioning for the initial data distribution.\footnote{ The initial data distribution is not a mapping solution, only an initial assignment of the input data to the PEs.}
Block partitioning of the input graph often leads to many inter-PE edges even before coarsening,
in particular for complex networks.
One reason for that is the high-degree nodes of complex networks, also known as hubs,
when they become halo vertices. When the number of hubs being halo vertices becomes
large enough, the scalability of \parhip is negatively affected.

\begin{figure}[bt]
  \centering
\includegraphics[scale=0.25]{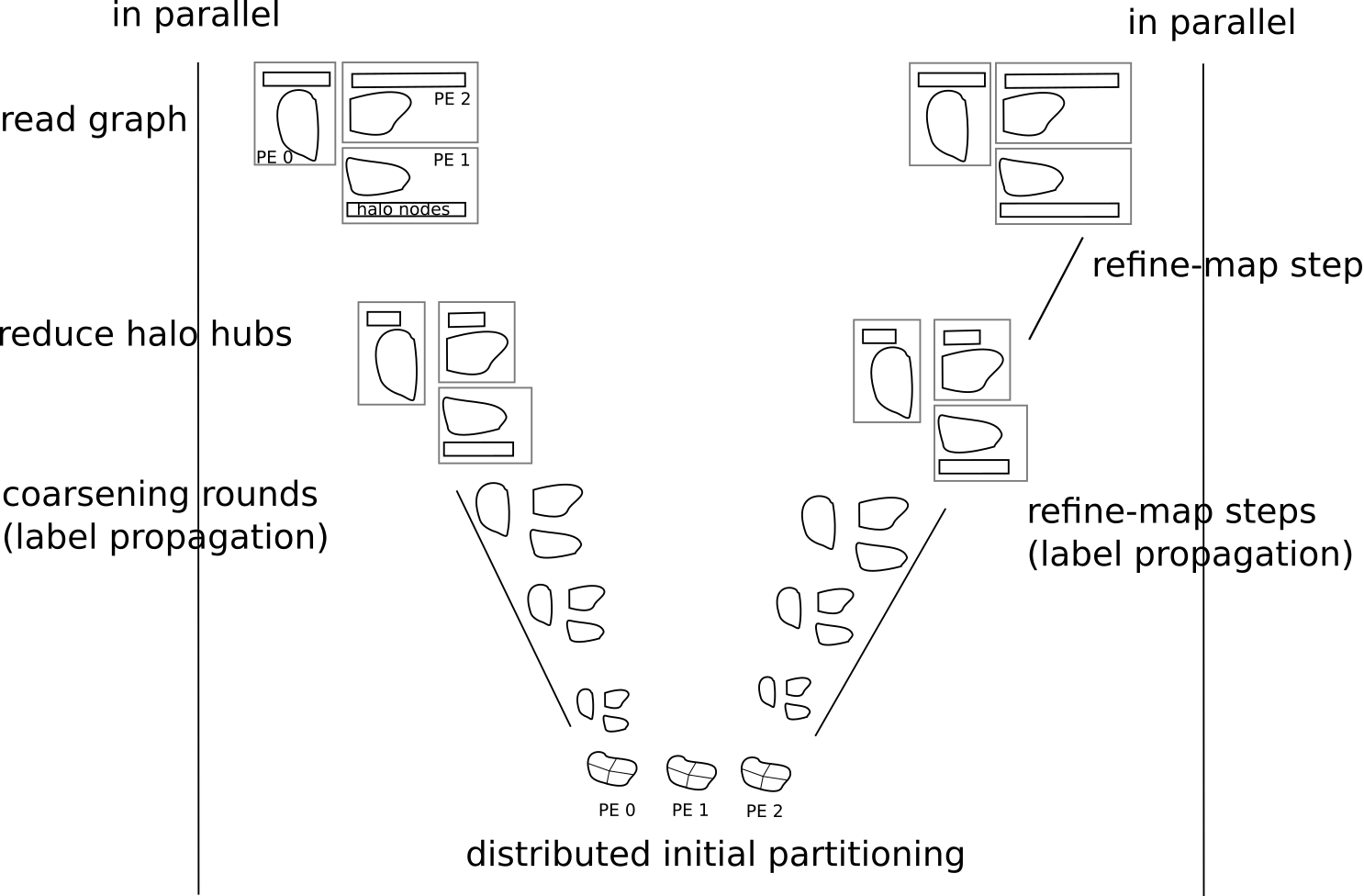} 
\caption{\label{fig:overall_algo}
  Schematic interpretation of the parallel mapping for
  an example with three PEs and a partition in four blocks.
  The algorithm performs one multi-level cycle with additional pre- and post-processing steps to
  avoid memory issues.
}
  \end{figure}

In our implementation we propose a simple correction for such
instances to avoid high memory usage:
first, all PEs globally identify halo nodes of high degree (halo hubs) and then
each PE temporarily removes edges connecting halo hubs with non-local nodes,
creating a reduced local graph. Then each PE runs \ouralgo on its reduced graph
and after completion each PE re-introduces the removed edges.
Once the one multilevel cycle is complete, we perform a few extra
rounds of parallel label propagation on the original graph
to compensate for quality losses due to the re-introduction of the removed edges.
This way we avoid the high memory consumption and save communication
during coarsening and initial 
partition steps without sacrificing much of the quality.
A schematic interpretation of the overall parallel algorithm is given in Figure~\ref{fig:overall_algo}.

\section{Experiments}
\label{sec:exp-setup}

\ifthenelse{\boolean{confversion}}
           {\budget{3.5}}
           {\budget{4.5}}

We perform experiments to evaluate the behavior
of \ouralgo on several graphs (see Table~\ref{tab:graphs}) downloaded
from SNAP~\cite{Leskovecxxxx} or generated
via KaGen~\cite{funke2017communication} and ParMAT~\cite{wsvr}. 
For disconnected graphs (in practice only some R-mat graphs)
we extract the largest connected component.
We implement \ouralgo in C++, using the \parhip graph API.
For performance experiments, we compare against MPI-based partitioning tools,
\parmetis 4.0.3, \parhip 3.10 (vanilla version configured to fastsocial) and \xpulp 0.3.
As mentioned in Sec.~\ref{sec:rel}, there are
no direct competitors for MPI-based mapping solutions so we use partitioning
with identity mapping. Often, identity mapping yields surprisingly good solutions,
since it benefits from spatial locality~\cite{Glantz2015c}.
Experiments were conducted on our local cluster\footnote{\url{https://www2.hu-berlin.de/macsy/technical-overview.html}}
or the HLRN\footnote{\url{https://www.hlrn.de/}} cluster, Lise, in Berlin.
Our local cluster contains 16 Linux machines, each
equipped with an Intel Xeon X7460 CPU (2 sockets, 12 cores each), and 192 GB RAM.
In Lise, each compute node has two Intel Cascade Lake Platinum 9242 CPUs with 384 GB RAM 
and 96 cores. Unless otherwise specified, we use one MPI process
per compute node.
%
%
%
We use the default settings for the competing algorithms,
and similar build settings among all codes.\footnote{On our local cluster: g++ 8.3.1
  compiler with -O2 flags and the mpich 3.2 MPI library. On Lise: g++ 9.2 compiler with -O2 flags and openmpi 3.1.5.} 
For all experiments, we report geometric mean results relative to \ouralgo over all graphs (of Table~\ref{tab:graphs}). For each graph we repeat
the experiment three times and we set the imbalance tolerance to 3\%, one of the values
used in~\cite{Walshaw-MPDD-07} and in \parmetis.
To ensure reproducibility, all experiments were managed by
SimexPal~\cite{angriman2019guidelines}. Our code and the experimental
pipeline can be found at~\url{https://github.com/hu-macsy/KaHIP}.

\begin{table*}
    \centering
    \caption{
      16 (undirected) graphs used in our experiments. Columns correspond to:
      name, type, number of vertices, number of edges, degree (average and max).
    }
    \label{tab:graphs}
\begin{scriptsize}
\begin{tabular}{|l|r|r|r|r|r|}
  
\hline
  Network &Type &$|V|$ &$|E|$ & $d_{avg}$ &$d_{max}$ \\
\hline
 
coPapersCiteseer 	&  \real & \numprint{434102}	& \numprint{16036720} 	& \numprint{73.8} & \numprint{1188} \\
eu-2005 &  & \numprint{862664} & \numprint{16138468} & \numprint{37.4} & \numprint{68963} \\
as-skitter      &   & \numprint{1696415}   & \numprint{11095298}   & \numprint{13} & \numprint{35455} \\
orkut     &   & \numprint{3072441}   & \numprint{117184899}   & \numprint{76.2} & \numprint{33313} \\
dbpedia  & & 	\numprint{18265512} & \numprint{136535446} & \numprint{14.9} & \numprint{612308} \\  
friendster &   & \numprint{65608366}    &	\numprint{1806067135} & \numprint{55} & \numprint{612308} \\ 	
twitter   &    & \numprint{52515193} &  \numprint{1963197641} & \numprint{74.7} & \numprint{3691240} \\
\hline


r-mat         ($\times$3) &  \rmat    & $2^{22} - 2^{24}$    &   $2^{27} - 2^{29}$    &  \numprint{40} & \numprint{18484}-\numprint{63345} \\
barabasi-albert ($\times$3)      &  \ba    & $2^{23} - 2^{27}$    &   $2^{26} - 2^{32}$    &  \numprint{32} & \numprint{19905}-\numprint{40509} \\
random-hyperbolic ($\times$3)     &  \rhg    & $2^{25} - 2^{29}$    &   $2^{29} - 2^{33}$    & \numprint{16} & \numprint{83645}-\numprint{200165} \\
\hline

\end{tabular}
\end{scriptsize}


\end{table*}

\subsection{Parallel Performance}
\label{exp:parallel-scalability}

\begin{figure*}
\centering
\begin{subfigure}{\textwidth}
  \centering
    \includegraphics[scale=0.27]{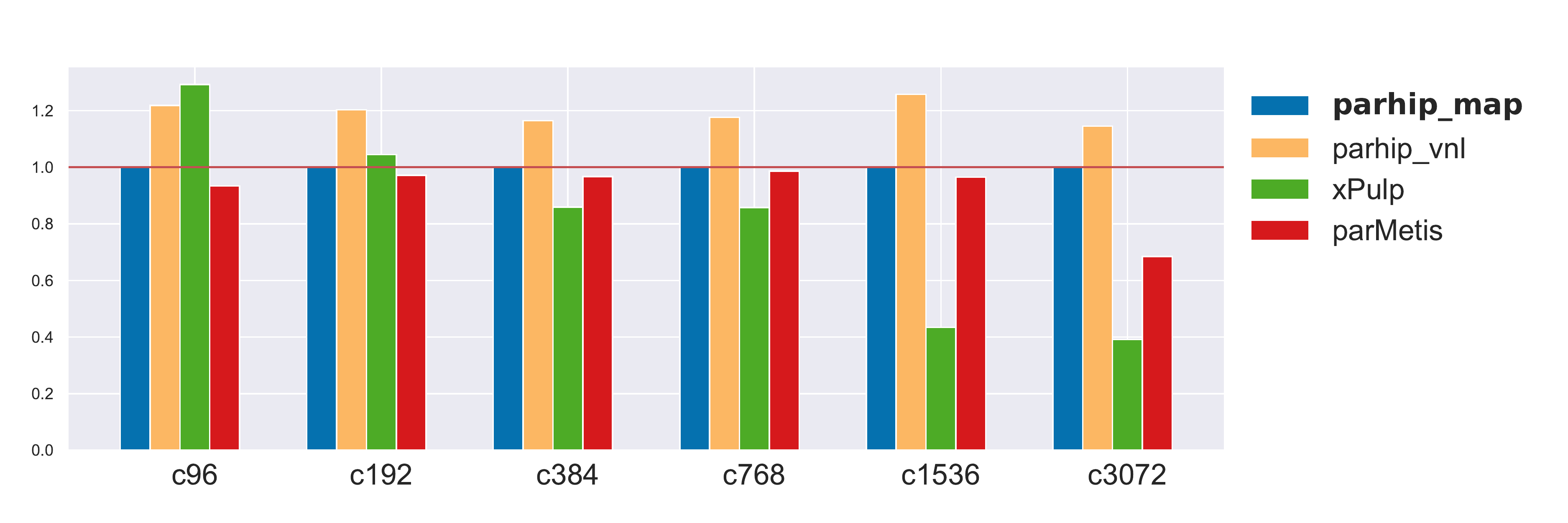}
  \caption{
      \label{fig:lise:var_pes:run}
 Results for scaling PEs and constant number of blocks$=1536$ on Lise.}
\end{subfigure}
\begin{subfigure}{\textwidth}
\centering
  \includegraphics[scale=0.27]{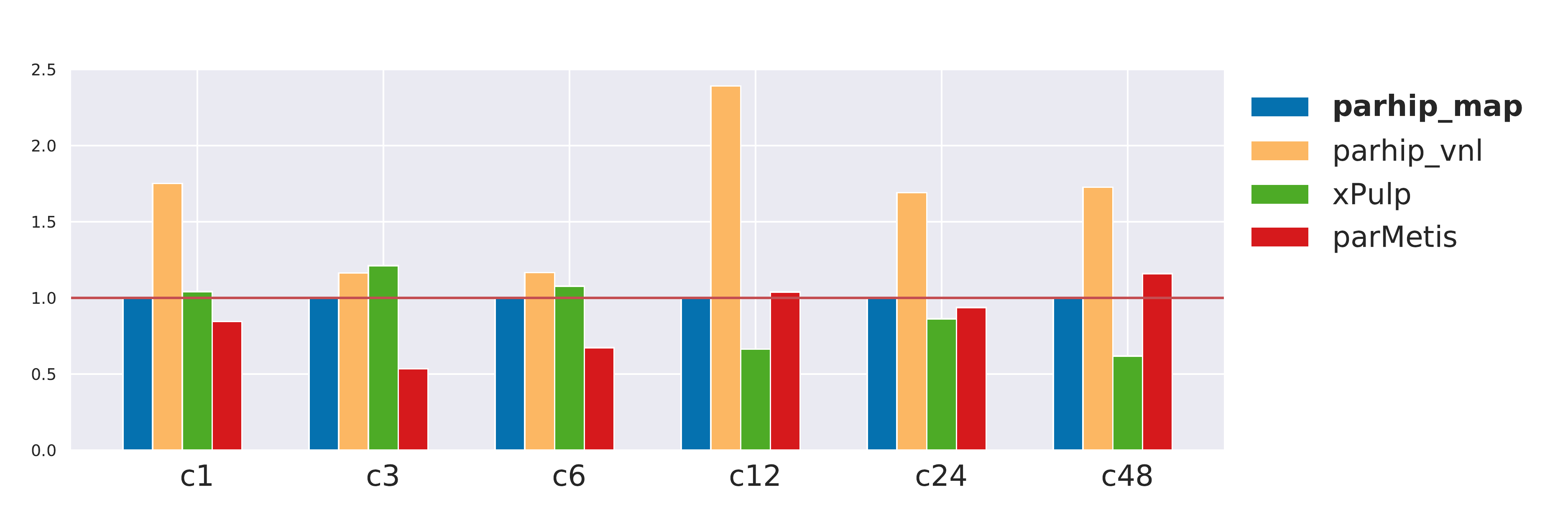}
  \caption{
      \label{fig:var_pes:run}
 Results for scaling PEs and constant number of blocks$=384$ on our local cluster.}
\end{subfigure}
\begin{subfigure}{\textwidth}
\centering
\includegraphics[scale=0.27]{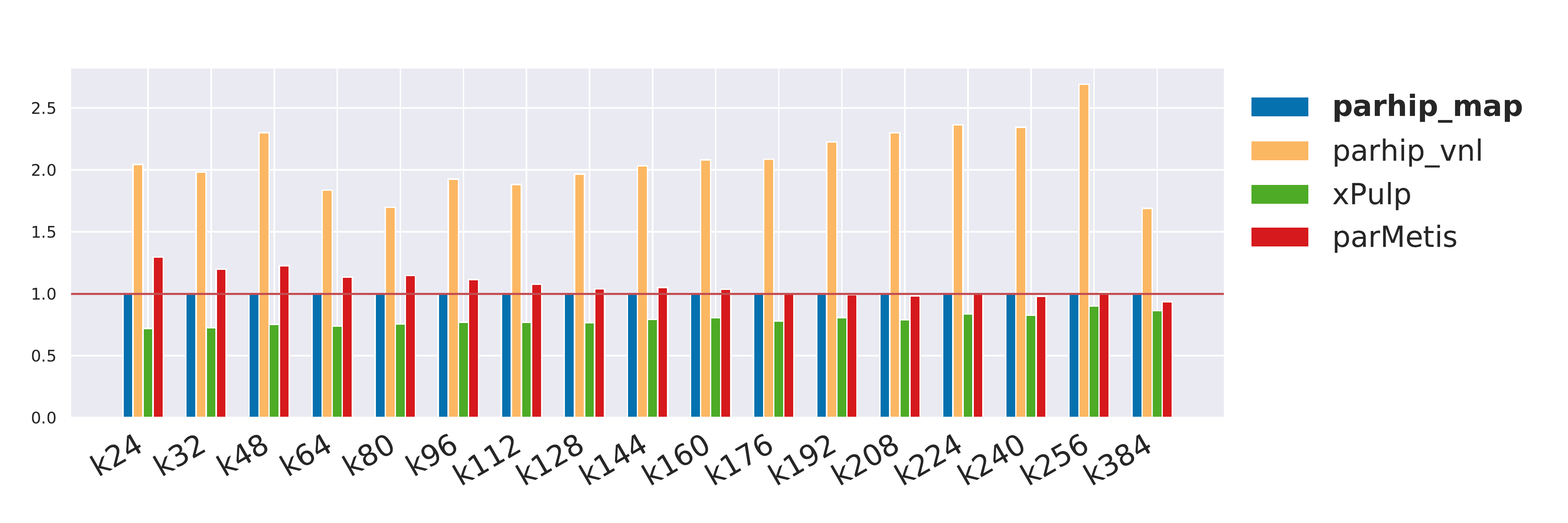}
\caption{
  \label{fig:var_blocks:run}
  Results for scaling number of blocks and constant number of PEs$=24$ on our local cluster.}
\end{subfigure}
  \caption{
    \label{fig:q:var_block}
          {\small Relative running times for different scaling experiments on various clusters.}
  }
\end{figure*}

We first evaluate the scalability behavior of \ouralgo
in a massively parallel setting of up to \numprint{3072} PEs on Lise.
In Figure~\ref{fig:lise:var_pes:run} we report running times for all
parallel tools relative to \ouralgo. The results indicate that
\ouralgo exhibits a good scaling behavior. Compared to the fastest tools
\ie,~\xpulp and \parmetis, \ouralgo is on average only 18\% and 9\% slower, respectively.
The high speed of \xpulp is not surprising since it is designed
to explicitly favor speed over quality.
It is important to note that Figure~\ref{fig:lise:var_pes:run}
depicts aggregated results that may hide
out-of-memory or timeout issues. After examining the failing rates, we
observed that \ouralgo has the smallest failing rate (17\%), followed by \xpulp, \parhip
and \parmetis with failing rates of 42\%, 62\% and 65\% respectively. 
To reflect a fairer comparison, we also include scalability experiments on our local
cluster\footnote{Here, we use one MPI process per core.},
where we observe a slightly better scaling behavior for \ouralgo and similar trends
for the other competitors (see Figure~\ref{fig:var_pes:run}).
It is noteworthy to report that \ouralgo
is the only tool that runs successfully for the twitter graph. Precisely,
our algorithm maps the twitter graph (a graph in the billion-scale range) into $384$ blocks
on 48 PEs of our local cluster in less than 6 minutes. 
\parmetis and \parhip failed for almost all Barabasi-Albert
and R-mat graphs, probably due to the highly irregular degree distribution
of these graphs, leading to memory issues.
Moreover, in Figure~\ref{fig:var_blocks:run}
we report scaling results for an increasing number of blocks.
We clearly see that \ouralgo is on average $2\times $ faster than \parhip,
slightly faster than \parmetis and only about $0.7\times$ slower than \xpulp.

\begin{figure*}
\centering
\includegraphics[scale=0.29]{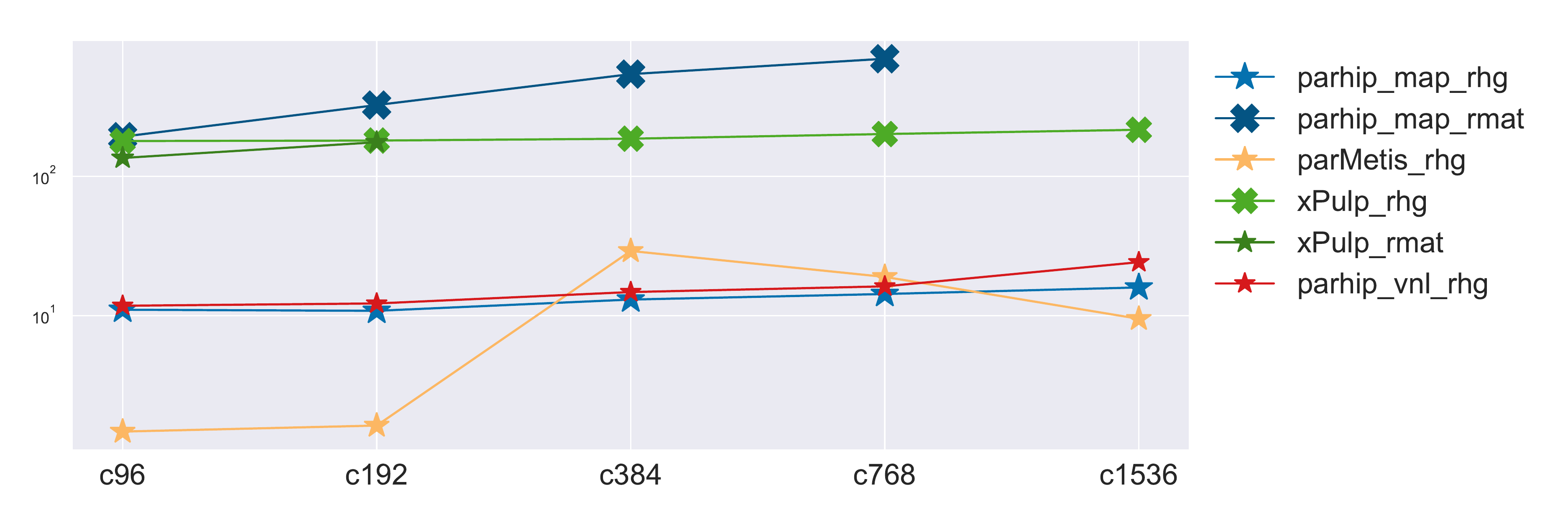}
  \caption{
    \label{weak_sca}
          {\small Absolute running times for weak scaling experiments on Lise (logarithmic scale).}
  }
\end{figure*}

We also perform weak scaling experiments on Lise,
for R-mat and random hyperbolic graphs of different sizes,
for up to \numprint{768} and \numprint{1536} PEs,
respectively.
For the experiment, we double the number of vertices as PEs double.
The number of blocks is equal to the
number of PEs used in the run
and missing inputs are due to failing runs.
In Figure~\ref{weak_sca}, we see that \ouralgo has
a similar scaling behavior as \xpulp while the latter
is faster as already observed from strong scaling.

\subsection{ Quality Results }
\label{exp:qap}

\begin{figure*}
\centering
\includegraphics[scale=0.32]{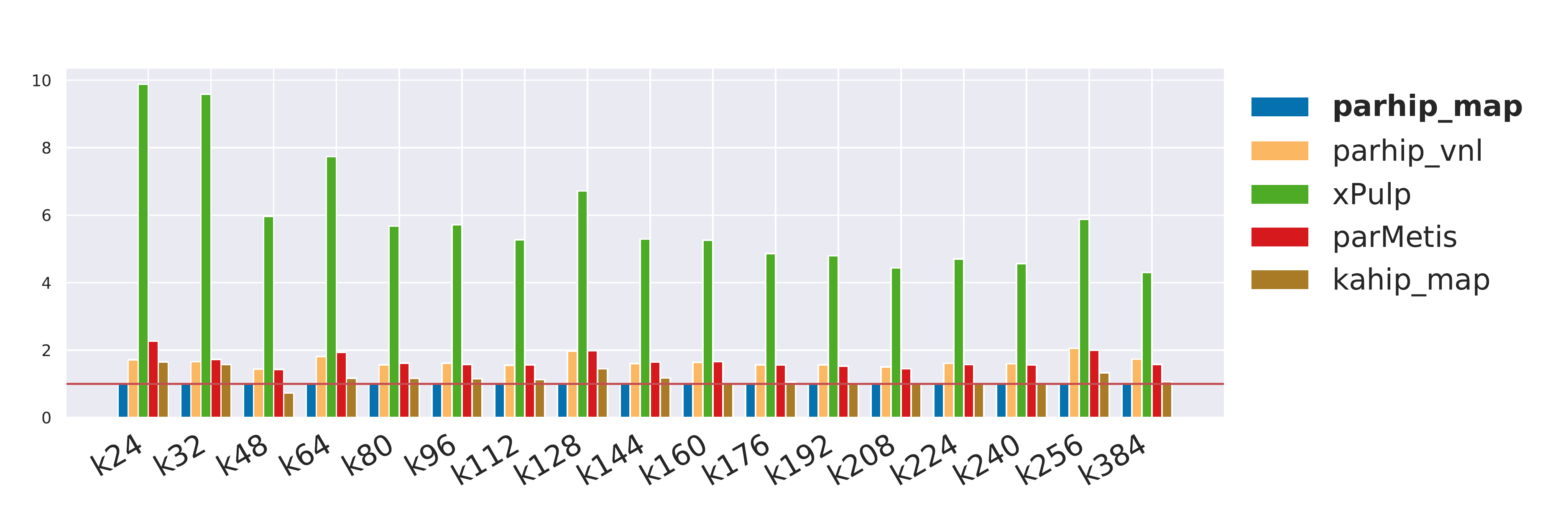}
  \caption{
    \label{var_block:qap}
          {\small Relative $\cc$ results for $24$ PEs (for the parallel tools).
          Lower is better.}
  }
\end{figure*}

To evaluate the solution quality, we use the objectives $\cc$ and edgecut and run
all parallel competitors as well as a sequential mapping approach from \kahip, named here \mapkahip
\footnote{Here, we set \mapkahip to the fastsocial configuration.}, known to produce
mapping results of high quality~\cite{faraj_et_al:LIPIcs:2020:12078}.
In Figure~\ref{var_block:qap}, we present relative $\cc$ results
for a constant number of PEs and an increasing number of
blocks, since the latter often affects quality.
For edgecut, we report results directly in the text, due to space limitations.
Figure~\ref{var_block:qap} shows that \ouralgo achieves consistently
the best $\cc$ results compared to all other parallel approaches.
More precisely, we are, on average, at least $ 4 \times$ better than \xpulp, 
62\% better than \parhip, and 70\% better than \parmetis.
Regarding edgecut, we are only
10\% worse than the best competitor (\parhip), 5\% than \parmetis,
but $2.5\times$ better than \xpulp.
Those results are surprisingly good for
our algorithm -- given the fact that we do not optimize for edgecut, as the competitors do.
Regarding the sequential baseline, \mapkahip, we even achieve better quality (\ouralgo is 10\% better) 
and we are $30\times$ faster using 24 PEs (as to 1 for \mapkahip).
Note that \mapkahip, also fails 
to finish in time or has memory issues in many cases.
Finally we should report that all tools occasionally fail to adhere to the balance constraint of 3\% but do not largely overpass it either.

\section{Conclusions}
\budget{0.5}

In this work we propose a fully parallel mapping algorithm for distributed-memory systems.
Our algorithm is an integrated solution \ie, it addresses the partitioning and mapping
problems simultaneously. We target hierachical systems
and encode the hierarchy with a concise representation using bit-labels.
Our approach exploits the above representation
and uses parallel label propagation to devise a fast refinement process.
As far as we know, this is the first parallel mapping algorithm for
distributed-memory systems with a publicly available implementation (within \parhip).
Given the experimental results, our algorithm offers the best trade-off between
mapping quality and speed compared to other MPI-based approaches. 
For future work we would like to integrate more scalable
initial partitioning techniques (like the one proposed in~\cite{faraj_et_al:LIPIcs:2020:12078})
to improve the performance of our current implementation.

\paragraph*{Acknowledgements.}
{\small 
This work was partially supported by the
North-German Supercomputing Alliance (HLRN). We also
thank our colleague Fabian Brandt-Tumescheit for his technical support regarding our group's cluster.}

\bibliographystyle{splncs04}

\bibliography{paper-dmap}
\budget{2.5}

\end{document}